\documentclass[sigconf]{acmart}
\AtBeginDocument{%
  }

\setcopyright{acmlicensed}
\copyrightyear{2026}
\acmYear{2026}
\setcopyright{cc}
\setcctype{by}
\acmConference[ACM Sustainability Week Companion '26]{ACM Sustainability Week 2026}{June 22--25, 2026}{Banff, AB, Canada}
\acmBooktitle{ACM Sustainability Week 2026 (ACM Sustainability Week Companion '26), June 22--25, 2026, Banff, AB, Canada}
\acmDOI{10.1145/3765611.3815432}
\acmISBN{979-8-4007-2199-1/2026/06}

\usepackage{listings}
\usepackage{xcolor}
\makeatletter
\renewcommand{\@fnsymbol}[1]{%
  \ifcase#1\or
    \ensuremath{*}\or
    \ensuremath{**}\or
    \ensuremath{\dagger\dagger}%
  \else
    \@ctrerr
  \fi
}
\makeatother

\usepackage{algorithmic}
\usepackage{algorithm}
\begin{document}

\title[BuilDyn]{BuilDyn: Excitation-Driven Data Generation for Building Thermal Dynamics Modeling and Control}


\author{Felix Koch}
\authornote{Also with Karlsruhe Institute of Technology.}
\affiliation{%
  \institution{Technical University of Applied Sciences Rosenheim}
  \city{Rosenheim}
  \country{Germany}}
\email{felix.koch@th-rosenheim.de}

\author{Thomas Krug}
\authornotemark[1]
\affiliation{%
  \institution{Technical University of Applied Sciences Rosenheim}
  \city{Rosenheim}
  \country{Germany}}
\email{thomas.krug@th-rosenheim.de}

\author{Fabian Raisch}
\authornote{Main affiliation at Technical University of Applied Sciences Rosenheim; doctoral candidate at Technical University of Munich (cooperative doctorate with Technical University of Applied Sciences Rosenheim).}
\affiliation{%
  \institution{Technical University of Applied Sciences Rosenheim}
  \city{Rosenheim}
  \country{Germany}}
\affiliation{%
\institution{Technical University of Munich}
  \city{Munich}
  \country{Germany}}

\author{Benjamin Schäfer}
\affiliation{%
  \institution{Karlsruhe Institute of Technology}
  \city{Karlsruhe}
  \country{Germany}}
\email{benjamin.schaefer@kit.edu}

\author{Benjamin Tischler}
\affiliation{%
  \institution{Technical University of Applied Sciences Rosenheim}
  \city{Rosenheim}
  \country{Germany}}
\email{benjamin.tischler@th-rosenheim.de}

%\authornotetext[1]{Also with University X.}

\renewcommand{\shortauthors}{Koch et al.}

\begin{abstract}

Machine learning (ML) is increasingly used for data-driven modeling of buildings to enable downstream tasks such as fault detection and diagnosis, and energy-efficient control. While recent work improves generalization across building characteristics, weather, and occupancy, generalization also depends on sufficient exploration of the control-driven system state space. Existing real-world datasets and simulation environments predominantly reflect stationary operation under fixed control policies, resulting in limited excitation and reduced robustness to unseen operating conditions.

This paper introduces BuilDyn, a package based on BuilDa that enables customizable excitation strategies for control-oriented data generation. BuilDyn further supports sampling from representative building distributions and provides a Python interface for easy integration into machine learning pipelines. We demonstrate the benefits of BuilDyn by comparing the performance of data-driven ML models trained on non-excited and excited data for one building. With BuilDyn, we hope to advance scalable control-oriented modeling and support future directions such as transfer learning and building-specific foundation models.

\end{abstract}




\received{20 February 2007}
\received[revised]{12 March 2009}
\received[accepted]{5 June 2009}

\maketitle

%\newpage
\section{Introduction}
\label{sec:intro}

Building operations account for roughly 30\% of global energy consumption \cite{eea2023DecarbonisingHeatingCooling}. 
Consequently, reducing energy demand through fault detection and diagnosis (FDD), as well as energy-efficient control, 
applications due to its scalability across diverse buildings and its ability to learn complex system behavior from data. 
In several studies, Machine Learning (ML) has demonstrated superior performance to conventional methods 
\cite{choi2023performance, mulayim2024tsfmbuildingsrevolutionize}. ML-based modeling has demonstrated the option of generalization through transfer learning, 
thereby addressing data availability as in \cite{raisch2025gentlgeneraltransferlearning, pinto2022sharingiscaring, neural2026}, 
and has been shown to have the best performance in continuous learning frameworks \cite{RAISCH2026116868}. However, ML methods have a key limitation: 
poor performance on out-of-distribution data \cite{countaRL2026}. 
This makes data generated through state-space exploration essential, as models trained on insufficiently excited operating conditions fail to generalize to unseen system dynamics and control regimes. Excitation denotes the deliberate stimulation of control inputs to actively explore the state-space.

A similar challenge appears in conventional data-driven modeling for buildings based on Resistance-Capacitance (RC) gray-box models, where low-quality data leads to poor parameter 
estimation \cite{SERASINGHE2024114123}. To address this issue, studies employ excitation strategies to enhance data richness and improve 
model accuracy. For example, Madsen and Schultz \cite{bacher_identifying_2011, madsen1993short} proposed a pseudo-random binary sequence 
(PRBS) with both short and long time constants to capture air and envelope dynamics. Other works similarly use tailored excitation signals to 
improve parameter estimation \cite{100c2e3b8bb24a83bab3ae769708cec8, KNUDSEN2021117227, SARTORI2023109149}. However, manipulating the operation of a building to collect high-quality data is often not possible in real buildings due to occupancy constraints \cite{HARB2016199}.

An alternative for both conventional data-driven and ML-based approaches is the use of simulated datasets. For example, \cite{SARTORI2023109149} provides data 
from six real-world buildings, including dedicated excitation periods that enable efficient ML-based modeling and control, as demonstrated in 
\cite{yang2021experiment}. However, such datasets are typically designed for specific use cases and have restricted large-scale applicability. 
Large-scale datasets of real-world or simulated buildings, such as \cite{berkes2025hotdataset, luo2022ecobee, pullinger2021ideal}, partially 
address this limitation. Such datasets are particularly valuable for transfer learning applications, which use large-scale datasets to train ML models that can be later fine-tuned to real buildings with limited data \cite{dou2025transfer}%, neural2026}.
Nevertheless, these large-scale datasets predominantly reflect operations constrained by existing control policies and therefore lack sufficient excitation, limiting their suitability for learning robust models for FDD and control.

A third option is to use data generators or dedicated simulation environments. These range from complex white-box models based on EnergyPlus \cite{crawley2001energyplus}, Modelica \cite{Wetter014buildinglib}, or TRNSYS \cite{Klein2017TRNSYS18} to more user-friendly environments that simplify control and benchmarking, such as Sinergym \cite{Sinergym2021} or BoPTEST \cite{blum2021boptest}.
White-box models offer the benefit of a complete physical record of the building dynamics, but they are computationally expensive and require significant expertise to set up and operate. 
More user-friendly environments provide easier access to building simulations, but often rely on a limited number of fixed building models and other constraints, which may not capture the diversity of real-world buildings.
Krug et al. \cite{krug2025builda2} combine the benefits of high flexibility and ease of use by providing an FMU-based simulation environment that can be customized to different building envelopes, locations, and occupancy profiles. 
So far, this tool lacks dedicated excitation strategies that enable control-oriented modeling.

To address this gap, we investigate excitations for buildings to generate more robust data for training ML-based building models using three different excitation strategies. We show how such strategies explore simulated building state-spaces more thoroughly than in controlled buildings. Based on these excitation strategies, we show that data generated with such excitations can improve the predictive accuracy of ML models across a wider range of state spaces and dynamics. Additionally, we introduce the Python package \textit{BuilDyn}. The package allows for the implementation of custom excitation strategies for building simulations based on BuilDa \cite{krug2025builda2} to generate rich datasets to train and evaluate ML models for control. \textit{BuilDyn} also extends the original framework by enabling sampling from representative distributions of building domains (e.g., buildings constructed between 1980 and 2000 in a specific country), combined with functionalities for easier use.

\iffalse
To address this gap, we introduce BuilDyn, an extension of \cite{krug2025builda2} that enables configurable excitation strategies for building simulations. \textit{BuilDyn} allows customizable excitation strategies for building simulations, enabling the generation of rich datasets for training and evaluating ML models for FDD and control. 
In the experiments section, we demonstrate the benefits of \textit{BuilDyn} by comparing modeling performance on non-excited and excited data for one building. 
% fas fehlt hier nnoch?
In addition, \textit{BuilDyn} extends the original framework in two ways. First, it enables sampling from representative distributions of building domains (e.g., buildings constructed between 1980 and 2000 in a specific country), allowing users to generate populations of buildings with varying parameters, weather conditions, and occupancy profiles. Second, \textit{BuilDyn} is provided as a Python package, simplifying integration into ML pipelines and enabling reproducible benchmarking.
\fi

The remainder of this paper is structured as follows. Section \ref{sec:simulation} introduces \textit{BuilDyn}. Section \ref{sec:applications} evaluates excitation strategies from \textit{BuilDyn} on a building modeling task and compares excited and non-excited data. Finally, Section \ref{sec:conclusion} concludes the paper and discusses future applications.

\section{BuilDyn}
\label{sec:simulation}

%In this section, we introduce \textit{BuilDyn}, a Python package designed for the manipulation and simulation of Modelica-based \cite{mattsson1997modelica} FMU models, and dynamically add excitation patterns. The package is inspired by and built on BuilDa \cite{krug2025builda2}, a framework for generating large-scale, realistic building data for ML research. For an in-depth description of the functionalities of BuilDa and its configurable FMU, we refer to \cite{krug2025builda2}. For a detailed description regarding BuilDa and BuilDyn, we refer to the appendix.

\begin{figure}
  \centering
  \includegraphics[width=\linewidth]{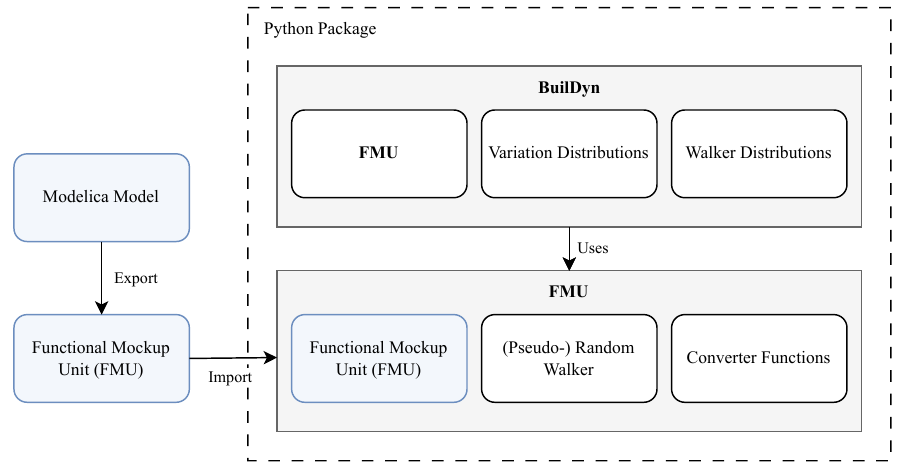}
  \caption{Structure of the \textit{BuilDyn} Python package.}
  \label{fig:fmux.pdf}
\end{figure}

In this section, we introduce \textit{BuilDyn}, a Python package that extends BuilDa \cite{krug2025builda2}. BuilDa provides a framework for generating large-scale, realistic building data, while \textit{BuilDyn} adds functionality to dynamically manipulate control inputs and generate excited trajectories for control-oriented modeling. For an in-depth description of BuilDa and its configurable FMU, we refer to \cite{krug2025builda2}; additional details on the relationship between BuilDa and \textit{BuilDyn} are provided in the appendix.
% \vspace{2pt}
% \noindent
% The \textit{BuilDyn} package builds on the following key concepts:

\subsection{Architecture}

%\vspace{4pt}
\noindent The \textit{BuilDyn} package consists of two main components visualized in Figure \ref{fig:fmux.pdf}: The FMU and the BuilDyn class. The FMU class is a wrapper for the FMU API FmPy \cite{dassault2023fmpy}, providing an interface to simulate buildings. In \textit{BuilDyn}, the FMU of BuilDa is slotted in. However, \textit{BuilDyn} is also designed to hold other FMUs. The BuilDyn class, on the other hand, is a wrapper for the FMU class that enables the creation of FMU objects and FMU simulations from user-defined distributions. \textit{BuilDyn} and further tutorials are also available on \href{https://github.com/felixmkoch/buildyn}{GitHub}.   

\begin{figure*}
  \centering
  \includegraphics[width=\linewidth]{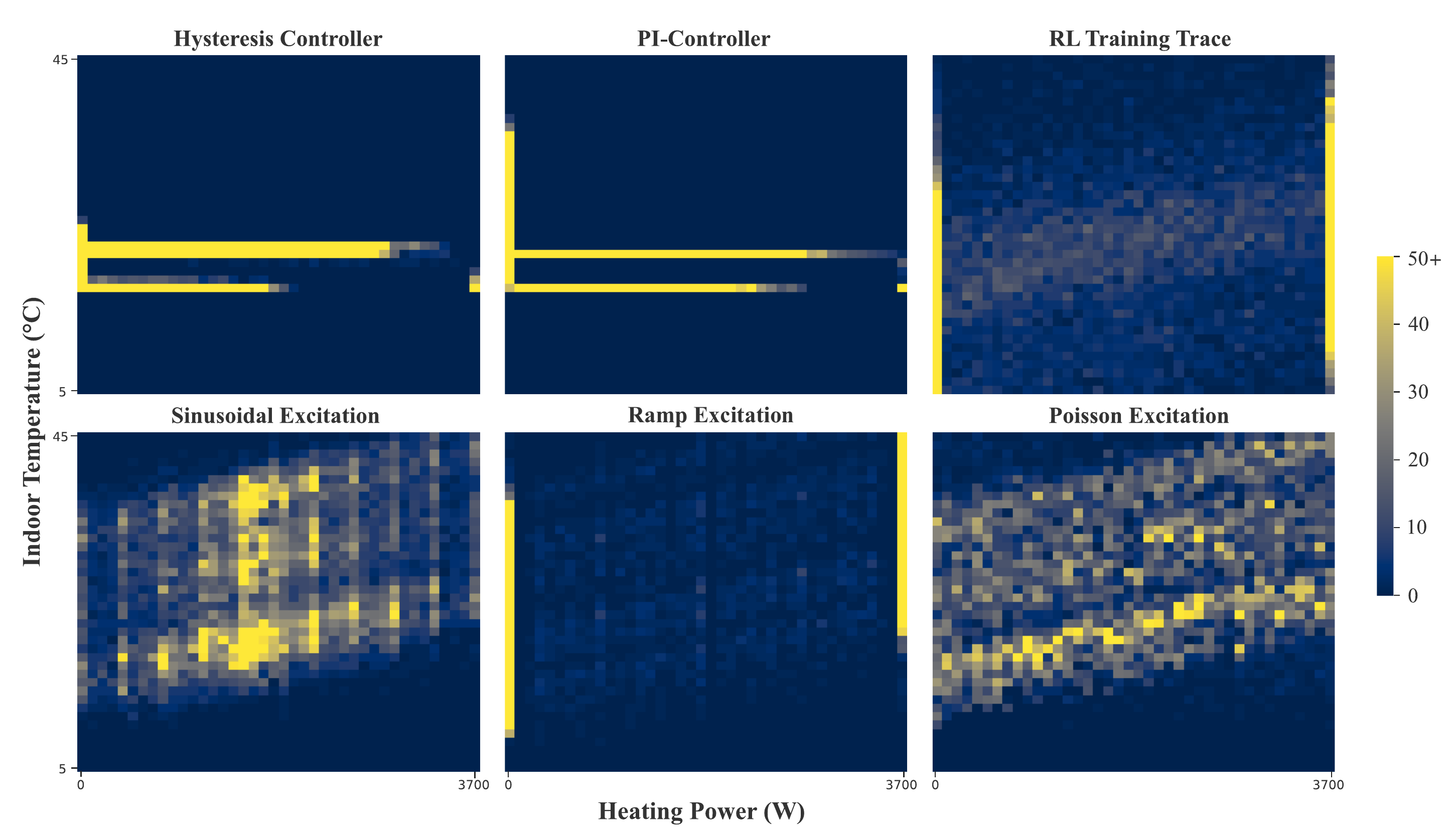}
  \vspace*{-0.7cm}
  \caption{State-Exploration of BuilDyn with different excitation/control strategies}
  \label{fig:buildyn_states.pdf}
\end{figure*}

\subsection{Key Concepts}
\label{sec:key_concepts}
\textit{BuilDyn} was designed incrementally on BuilDa around the following four key concepts/features.
%Apart from the extension through the excitation strategies (key concept F1), the \textit{BuilDyn} package provides other extensions, called key concepts. We explain each of them in the following:

\vspace{4pt}
\noindent \textbf{F1: Excitations.} Excitations, such as pseudo-random inputs or stochastic signals, are commonly used in system identification to explore different states of a system. \textit{BuilDyn} introduces the concept of walkers, which enable the user to define excitations for the dynamic system that trigger in configurable time intervals. These walkers enable more thorough state explorations of buildings compared to standard control methods. In particular, three complementary excitation strategies are employed: (i) a \textit{Poisson excitation}, which applies piecewise-constant heating levels for randomly sampled durations to create irregular step-like inputs; (ii) a \textit{sinusoidal excitation}, which chains sinusoidal segments with varying frequency and amplitude separated by short steady-state holds to mimic diverse periodic heating patterns; and (iii) a \textit{ramp excitation}, which repeatedly ramps linearly between minimum and maximum values with holds at the extrema to systematically probe transitional dynamics. Together, these walkers provide a mix of stochastic persistence, periodic variation, and structured transitions, improving coverage of the control-dependent state-space. For further details on the three strategies, see Appendix \ref{sec:appx:other_applications}.
%\textbf{F1: Excitations.} Excitations, such as pseudo-random inputs or stochastic excitations, are commonly used in system identification to explore different states of a system. \textit{BuilDyn} introduces the concept of walkers, which enable the user to define excitations for the dynamic system that trigger in configurable time intervals. These walkers enable more thorough state explorations of buildings compared to standard control methods.

\vspace{4pt}
\noindent \textbf{F2: User-Friendly FMUs.} \textit{BuilDyn} provides an FMU class that serves as an abstraction layer to the FmPy \cite{dassault2023fmpy} API. This abstraction provides users with a tool to simulate buildings and manipulate their envelope parameters, weather scenarios, and occupant profiles. Furthermore, \textit{BuilDyn}, as a Python package, is distributed on PyPI, enabling seamless integration of FMU functionality into pipelines that require building simulations.

\vspace{4pt}
\noindent \textbf{F3: Custom Variation Distributions.} Real-world building populations exhibit different distributions of envelope parameters and characteristics. With \textit{BuilDyn}, we enable users to custom define those distributions over building parameters to sample from. This enables users to reconstruct real-world building envelope distributions (for instance, centered on TABULA \cite{Loga2011-ws} archetypes) in their benchmarks.

\vspace{4pt}
\noindent \textbf{F4: Converter Functions.} Building parameters may require successive recalculations when modified. For instance, cooling-/heating-loads are dependent on the volume of a building. To allow users to define such chains of dependent updates, \textit{BuilDyn} adopts the idea of the converter layer functionality from \cite{krug2025builda2} but offers users full control over conversions through a dedicated Python interface.

\section{Experiments}
\label{sec:applications}
This section illustrates two experiments on different excitation strategies using the BuilDyn package. 
%can influence the performance of ML-models trained on the resulting data. 
In Section~\ref{sec:explore}, we compare different control and exploration strategies with respect to state-space coverage.
%we show that data simulated in buildings using standard controls can result in a narrow exploration of indoor air temperature and heating control signals. We then introduce three excitation strategies for buildings and show how they can widen the state-space explored during simulation. 
Subsequently, Section~\ref{sec:ddm} employs two of these strategies to generate data, which is then used to train an ML prediction model in the next step.
%Thereafter, in Section~\ref{sec:ddm}, we use two of these strategies to generate data that is used in a next step to train an ML model.
%In this example, we show that a broader state-space exploration of the building in the training data can improve the performance of a data-driven model for a building. Additional applications, experimental setups, and complementary results are further detailed in the Appendix. 

\subsection{State-exploration with excitations}
\label{sec:explore}
%Real-world buildings are typically occupied, requiring thermal comfort guarantees that are commonly enforced through controllers such as PI or hysteresis controllers. However, such controllers inherently restrict the range of explored building states, limiting the data available for tasks like system identification. Building simulations of such buildings offer a compelling alternative: since simulated environments do not need to fulfill thermal comfort constraints, they enable broader state-space exploration and, consequently, more accurate ML-based building models trained on sufficiently excited data. Nevertheless, existing work such as \cite{pinto2022sharingiscaring,RAISCH2026116868} often relies on controlled building simulations, resulting in similarly narrow state-space coverage during training. While this may suffice for tasks such as anomaly or fault detection, the resulting models are prone to extrapolation when deployed in contexts involving unseen behavior, for instance, as surrogate models within Model Predictive Control (MPC) or Reinforcement Learning (RL) frameworks. \textit{BuilDyn} addresses this limitation through the concept of \textit{walkers}, which enable systematic excitation of simulated buildings across a wide range of states.

%In this experiment, we compare two standard control variants with three excitation strategies obtained by BuilDyn. All three strategies are detailed in the appendix. 
%To illustrate the importance of this capability, we simulated one residential building under six distinct control and excitation strategies. 
In this experiment, we compare the three excitation strategies implemented in BuilDyn (see Section~\ref{sec:key_concepts}) with two standard control-based data generations and an RL training process.
For standard controls, we use a PI and a hysteresis controller, as these are typically used in reality. For demonstration, we use the standard pre-implemented building by \cite{krug2025builda2} in a heat control scenario. 
Figure~\ref{fig:buildyn_states.pdf} presents the resulting state-space coverage by counting and binning the appearing combinations of indoor temperature and heating signal for each 15-minute timestep over 1 year. %strategies (1) and (2) employ a PI and a hysteresis controller, respectively; strategy (3) corresponds to data collected during the training phase of an RL agent; and strategies (4), (5), and (6) represent customized \textit{BuilDyn} excitation approaches. 
The results demonstrate that the PI and hysteresis controllers only explore a narrow state band, as the system is only explored by driving disturbances, i.e., outdoor conditions and occupancy. The RL agent, on the contrary, explores a larger state space as part of its learning process. Consequently, ML models trained on data generated by these strategies more likely need to extrapolate when deployed in downstream tasks such as MPC- or RL-based control. In contrast, the three \textit{BuilDyn} excitation strategies exhibit substantially greater overlap with the RL trace, suggesting their superior suitability as training data for ML-based building models intended for control. In particular, the different excitation strategies cover different parts of the TL training trace in this concrete example.

\begin{figure*}
  \centering
  \includegraphics[width=\linewidth]{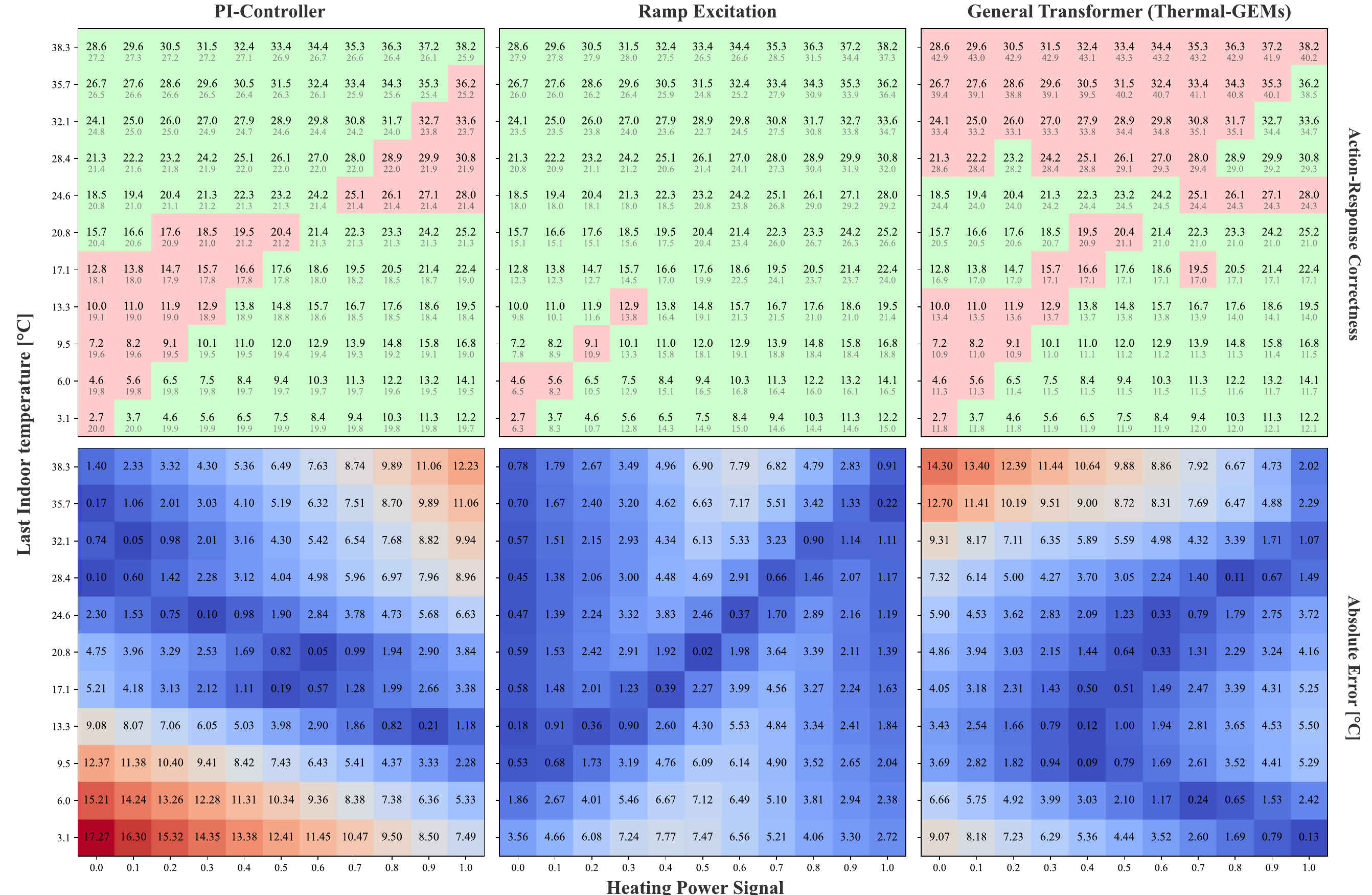}
  \caption{Top row: ARC of ML-Models with differently excited/controlled training data. Green background indicates correct response direction. The larger cell fonts show the ground truth, and the grey numbers show the predicted indoor temperature. Bottom row: Shows AE values with a better fit indicated by a blue color.
  }
  \label{fig:combined_evaluation_heatmaps.pdf}
\end{figure*}
\newpage
\subsection{Excitation effects on ML models}
\label{sec:ddm}

In this section, two of the strategies from Section~\ref{sec:explore} are employed to develop an ML–based prediction model for building thermal dynamics. The first model is trained on data generated by the PI-controller simulation with a heating setpoint of 22 °C, similar to \cite{RAISCH2026116868}. The second model uses data originating from BuilDyn using the ramp excitation. In both cases, a timestep of 900 s (15 min) is used for a 1-year simulation (January to December). For training, we use a transformer model \cite{vaswani2017attention}, following \cite{koch26thermalgems}. Additionally, we compare a third model from \cite{koch26thermalgems} that is trained on multiple controlled source buildings, which provides generalization capabilities across locations, occupancies, and building properties.

All three models are evaluated on February 1 of the second year. To obtain diverse initial thermal conditions, the simulation is first run throughout January of the second year, resulting in different indoor temperatures at 00:00 on February 1st\footnote{At 00:00, the outdoor temperature was 0.0°C, and there was no solar radiation.}. From this initialized state, a set of discrete control actions ranging from 0 (off) to 1 (full power) is applied. For each action, the resulting indoor temperature at the next time step (00:15) is recorded and compared to the corresponding prediction of the ML model. Performance is assessed using the \textit{absolute error} (AE) between prediction model and ground truth as well as the \textit{action-response correctness} (ARC). The ARC evaluates whether the model correctly captures the physical direction of the temperature change.

Figure \ref{fig:combined_evaluation_heatmaps.pdf} illustrates model behaviors across the different training strategies. %We consider three models: (left) trained on PI-controlled data, (middle) trained on ramp excitation data, and (right) a general transformer model trained on multiple single-zone buildings from BuilDa. 
Results show more green entries in the first row and blue entries in the second row for the excited scenario. The model trained on data generated within a PI-controlled building achieved 76\% ARC, while the model trained on data from buildings with ramp excitations achieved a 96\% ARC, which is the highest among all excitation strategies applied (complementary results are illustrated in Figure \ref{fig:combined_eval_heatmap_app}). The generalized transformer model we employed, based on \cite{koch26thermalgems}, only achieved a 54\% ARC, further motivating the use of data generated in excited building simulations for generalized data-driven models. This indicates that excitation-based modeling improves robustness to unseen states, yielding lower absolute errors and more accurate control responses. In contrast, models trained without excitation exhibit pronounced extrapolation errors, often reverting to values typical of the training distribution.

\section{Conclusion}
\label{sec:conclusion}
We present \textit{BuilDyn}, an open-source Python package for simulating and exciting buildings modeled with FMUs. Using an example building simulated in \textit{BuilDyn}, we highlight the necessity of excitation in data generation for ML-based building modeling. We show how excitations defined in \textit{BuilDyn} can help mitigate extrapolation behavior in ML-based models for buildings.
Limitations of this work include the extrapolation behavior of ML-based building models, which has only been investigated superficially. Seasonal effects and other covariates beyond the control signal and indoor temperature have not been considered. Although we observed similar patterns when evaluating the behavior of ML models at other times and in other seasons, broader evaluations are required. As transfer learning is an efficient method for modeling building thermal dynamics \cite{raisch2025gentlgeneraltransferlearning}, future work may focus on excitation strategies to improve the robustness of generalized models under extrapolation. 
\textit{BuilDyn} is envisioned as a community tool for benchmarking, manipulating, and exciting buildings, thereby contributing to the development of more robust generalized building models.

%\newpage
\balance
\bibliographystyle{ACM-Reference-Format}
\bibliography{refs}

\newpage
\appendix

\section{Extended BuilDyn Description}
\label{sec:appx:other_features}
BuilDyn is currently designed as a wrapper around BuilDa \cite{krug2025builda2}, also employing the BuilDa-specific FMU. In this part, we describe \textit{BuilDyn} and BuilDa in more detail.

\subsection{FMU Description}

The BuilDa simulation framework \cite{krug2025builda, krug2025builda2} enables the generation of high-fidelity operational data for residential buildings with systematically varied characteristics. The framework is implemented in Modelica and exported as a Functional Mock-up Unit (FMU), allowing users to configure building parameters directly in Python without modifying the underlying simulation model. The simulation model underlying BuilDa was validated in accordance with the ANSI/ASHRAE 140-2004 standard using the benchmark test cases TC600, TC900, TC600FF, and TC900FF.
The model represents single-family houses with detailed envelope components, orientation-dependent windows, and a controllable heat/cool source sized to account for transmission and ventilation losses. Building properties such as insulation level, thermal mass, window-to-wall ratio, and floor area can be varied, and simulations can be performed for different weather locations. In addition, BuilDa provides a stochastic occupancy generator that produces user-specific temperature setpoints, internal heat gains, and window-opening schedules for natural ventilation based on the number of occupants and behavioral variability. These features enable the systematic generation of building datasets capturing variations in building characteristics, weather conditions, and occupant-driven operation.

\subsection{Differences towards BuilDa}

Table \ref{tab:fmux_vs_builda} contrasts the main differences between BuilDa and \textit{BuilDyn}. As BuilDa is specifically designed for data generation, \textit{BuilDyn} simplifies the simulation process and its integration into existing projects. Moreover, it introduces new functionalities, such as customizable excitations and variation distributions. 
For the scope of this paper, we utilize the FMU provided by BuilDa, which models a single-family home in central Europe. However, \textit{BuilDyn} can be extended with additional FMUs modeling other building scenarios.

\begin{table}[H]
\centering
\caption{Feature comparison BuilDa vs BuilDyn. Parentheses indicate that featurerequires substential effort implemented in the corresponding Tool/Extension.}
\label{tab:fmux_vs_builda}
\begin{tabular}{l | c | c}
\toprule
    & \textbf{BuilDa} & \textbf{BuilDyn} \\
    \hline \hline
    \multicolumn{3}{l}{\textbf{Features}}\\
    \hline
    Variations for Buildings & \checkmark & \checkmark \\
    %\hline
    Parallel Simulation & \checkmark & \checkmark \\
    %\hline
    % Note Felix: Ist hier Customizable Excitations genug? Das ist ja auch in BuilDa gegeben und würde dann vielleicht keine große Contribution auf Seite FMUX bedeuten?
    Customizable Excitations & (\checkmark) & \checkmark \\
    %\hline
    Controllers with FMU Feedback & \checkmark & (\checkmark) \\
    %\hline
    %Demand-Driven Simulation & \(\times\) & \checkmark \\
    Data Generation & \checkmark & \checkmark \\
    %\hline
    Variation Distributions & \(\times\) & \checkmark \\
    \hline \hline
    \multicolumn{3}{l}{\textbf{Usability}}\\
    \hline
    Customizable Converter Functions & (\checkmark) & \checkmark \\
    %Provides Python API & \(\times\) & \checkmark \\
    %\hline
    %User-Friendly Interface & \(\times\) & \checkmark \\
    User-Friendly Python API & \(\times\) & \checkmark \\

    \bottomrule

\end{tabular}
\end{table}

\section{Extended BuilDyn Experiments}
\label{sec:appx:other_applications}
In this section, we present the use of \textit{BuilDyn} in two different ways. First, we provide example code showing an application of the package and also illustrate the influence of variations on both the heating behavior and the energy consumption of a building. 

\subsection{Example simulation of building variations}

We showcase \textit{BuilDyn} using an example in which we simulate 40 PI-controlled buildings with identical construction under a Central European weather file, while varying the wall U-values. For each building, the indoor temperature and the heating load are recorded. Listing \ref{listing:exmaple_code} illustrates how a single building instance is sampled and simulated using \textit{BuilDyn}. For our example, we executed the last line of the code 40 times.
\vspace{8pt}
\begin{lstlisting}[language=python, caption={Simplified example code to generate one variated building simulation in BuilDyn.}, label={listing:exmaple_code}]
fmu = get_builda_fmu(use_controller=True)
observables = ["temperature", "load"]

buildyn = BuilDyn(fmu, observables)
wall_dist = GaussDistribution(mu=0.8, sigma=0.5)
buildyn.add_variation_distribution({"UExt", wall_dist})

# Get the simulation of a building that is randomly sampled from the wall_dist.
data = buildyn.sample_one(stop_time=15_552_000)
\end{lstlisting}

\begin{figure}[H]
  \centering
  \includegraphics[width=\linewidth]{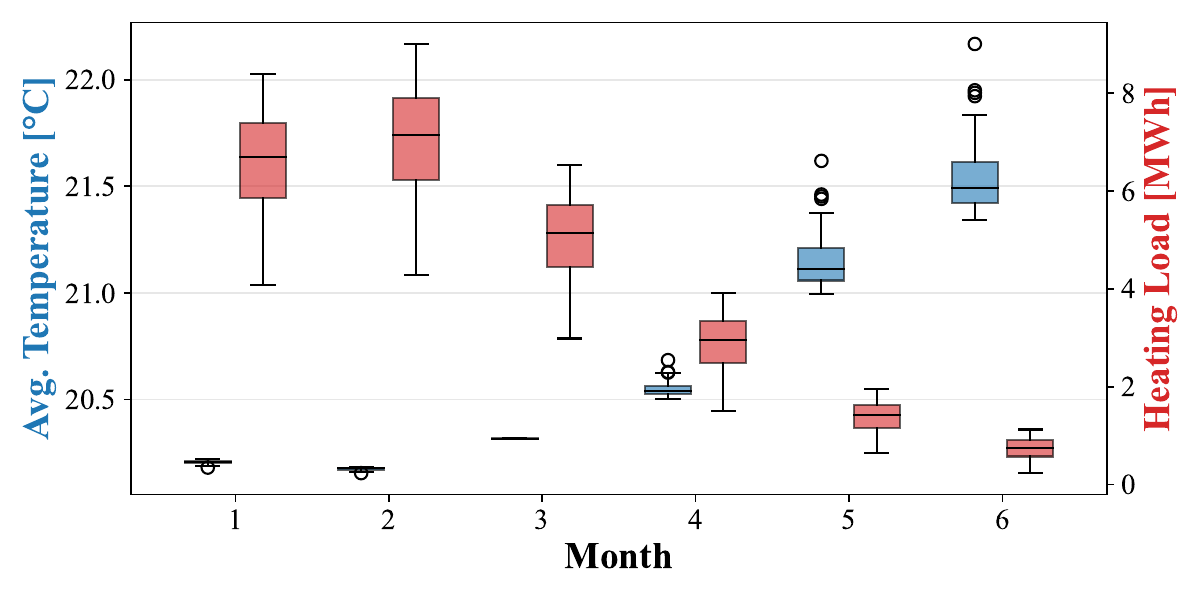}
  \caption{Heating load and indoor temperature distribution for 40 buildings.}
  \label{fig:load_dist.pdf}
\end{figure}

The result is visualized in Figure \ref{fig:load_dist.pdf}, showing the monthly distributions of indoor temperature and heating load across buildings as box plots over six months. The plot illustrates that the variation of a single parameter, in this case the wall U-value, can strongly influence the thermal behavior of the buildings. This example demonstrates how easily \textit{BuilDyn} can be configured to generate diverse residential building scenarios and produce substantial variability in thermal dynamics. 

\subsection{Customizable variation distributions}

In contrast to BuilDa, \textit{BuilDyn} enables both grid-based parameter variation and sampling from user-defined parameter distribution. In this example, we sample 40 buildings from custom distributions to show the difference. For that, we defined probability density functions (PDFs) over the sample space for the wall U-values and the floor area of a single-family home in Europe.

\begin{figure}[H]
  \centering
  \includegraphics[width=\linewidth]{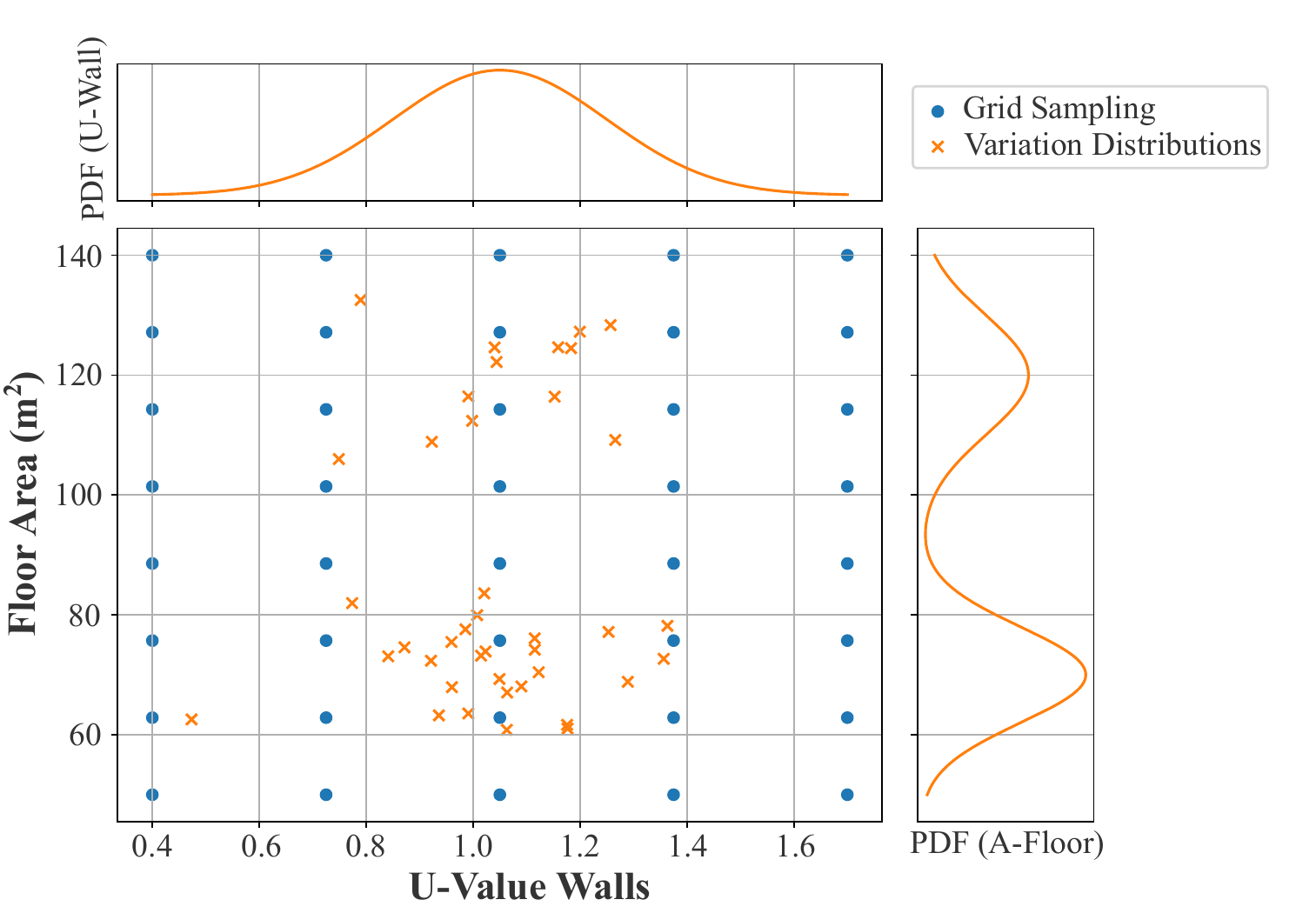}
  \caption{Sampling envelope parameter variations in \textit{BuilDyn} from a custom distribution, an unimodal for the wall U-values, and a bimodal for the floor area.}
  \label{fig:custom_variations}
\end{figure}

Figure \ref{fig:custom_variations} visualizes the difference between the grid-like and the variation-distribution-based sampling. Notably, the variation distributions in \textit{BuilDyn} are not restricted to Gaussian distributions, but can define all kinds of PDFs. Additionally, customizable variation distributions allow the reproduction of more realistic real-world building populations,
%that more closely resemble real-world distributions, 
which could support better benchmarking. These distributions are not restricted to continuous values over building envelope parameters, but also occupancy profiles or weather data files.

\iffalse
\section{Additional Results}
\label{sec:appx:additional_results}
\input{chapters/a3_additional_results}
\fi

\section{Experiment Details}
\label{sec:appx:hyperparameter}
This section contains additional information to fully reproduce the experiments from Section \ref{sec:applications}. All excitations and data can be reproduced using the \textit{BuilDyn} package. 

\subsection{ML Model Training}

For the transformer models, we adopted the hyperparameters from \cite{koch26thermalgems}. The architecture consists of three transformer encoder blocks followed by two dense layers serving as the decoder. To account for the single-building modeling task, rather than training on data from multiple buildings, the hidden size of both the transformer and dense layers was reduced from 192 to 96. For a fair comparison with \cite{koch26thermalgems}, we used a lookback of 96 (1 day) and a forecast horizon of 4; however, only the first forecasting step was considered in our experiments. Training was performed using the Adam optimizer \cite{kingma2014adam}. Early stopping was applied based on performance on a validation set, defined as the final month of the first year.

\subsection{Custom Excitation techniques}

Section \ref{sec:applications} showed three different excitation techniques to explore the building state-space better. This section showcases their implementation with pseudo-code. Notably, the respective distributions in the code are variable.

\vspace{4pt}
\noindent
\textbf{Poisson Excitation.} The idea behind the Poisson excitation is to input the same heating signal for a varying period of time. The number of repetitions is determined by sampling from a Poisson distribution.

\begin{lstlisting}[language=python, caption={Pseudo-code for the poisson excitation signals.}, label={listing:poisson_walker}]
result = []
while len(result) < num_steps:
    v = random.uniform(0, 1)  # heating signal
    k = random.poisson(lam)
    result.extend([v] * k)
return result[:num_steps]
\end{lstlisting}

\vspace{4pt}
\noindent
\textbf{Sinuoidal Excitation.} The sinusoidal walker generates an excitation signal by chaining sinusoidal segments of randomly sampled frequency and amplitude, separated by brief steady-state holds. This should simulate different kinds of heating patterns.

\begin{lstlisting}[language=python, caption={Pseudo-code for the sinusoidal excitation signals.}, label={listing:sinusoidal_walker}]
result, phase = [], 0.0
mid = 0.5
while len(result) < num_steps:
    freq = sample(freq_dist)
    amp  = sample(amp_dist)
    hold = sample(steady_dist)
    omega = 2 * pi / freq
    for _ in range(freq):
        v = clip(mid + 0.5 * amp * sin(phase))
        result.append(v)
        phase += omega
    result.extend([result[-1]] * hold)
return result[:num_steps]
\end{lstlisting}

\vspace{4pt}
\noindent
\textbf{Ramp Excitation.} This excitation strategy is repeatedly ramping linearly from a minimum to a maximum value and back, with brief steady-state holds at each peak and trough.

\begin{lstlisting}[language=python, caption={Pseudo-code for the ramp excitation signals.}, label={listing:ramp_walker}]
result = []
while len(result) < num_steps:
    freq = sample(freq_dist)
    step = 1 / (freq - 1)
    for i in range(freq):   # Ramp up
        result += [min + step * i] 
    # hold high
    result.extend([1] * sample(steady_dist))
    for i in range(1, freq-1):  # Ramp down
        result.extend([1 - step * i]) 
    # hold low
    result.extend([0] * sample(steady_dist))  
return result[:num_steps]
\end{lstlisting}

\begin{figure*}
  \centering
  \includegraphics[width=\linewidth]{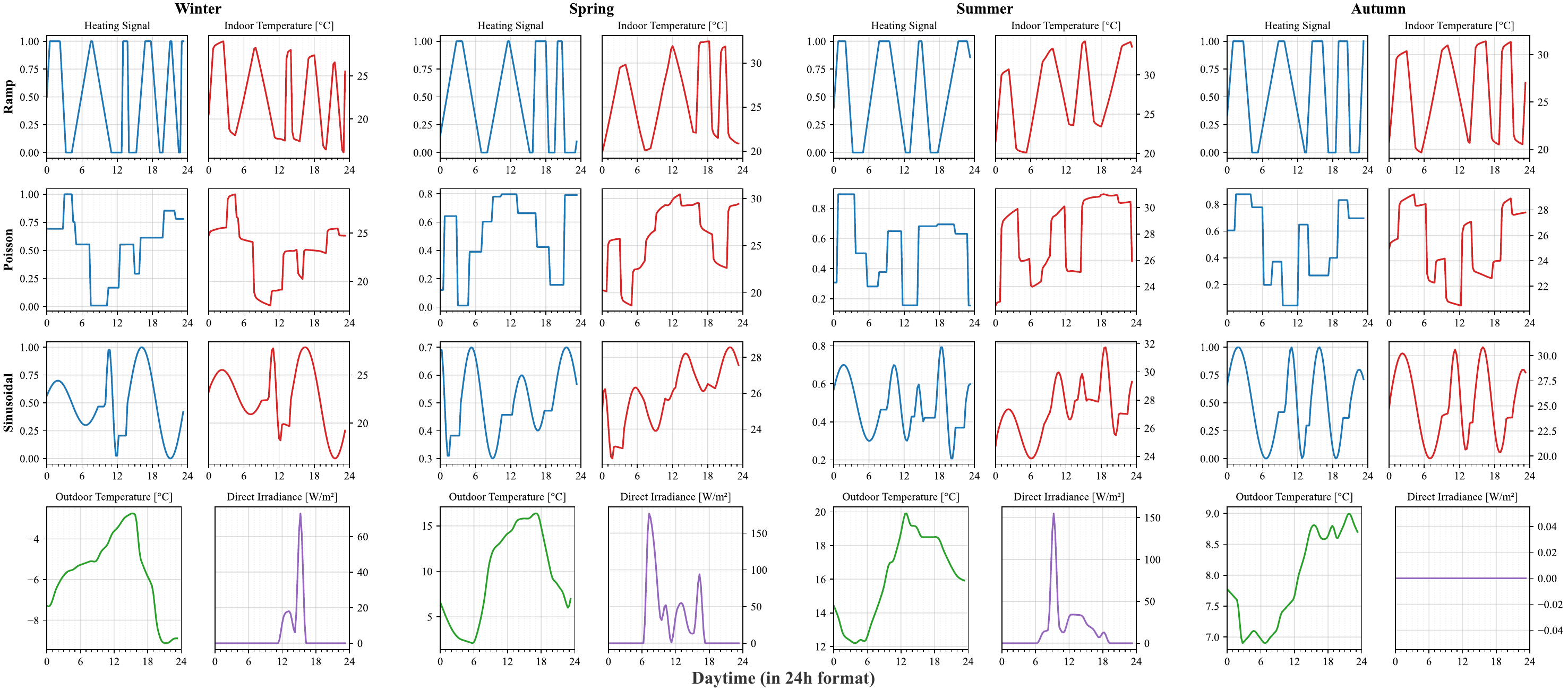}
  \caption{One day of excitation strategies that were used to create training data visualized for multiple seasons.}
  \label{fig:excitation_scenarios_app}
\end{figure*}

\begin{figure*}
  \centering
  \includegraphics[width=\linewidth]{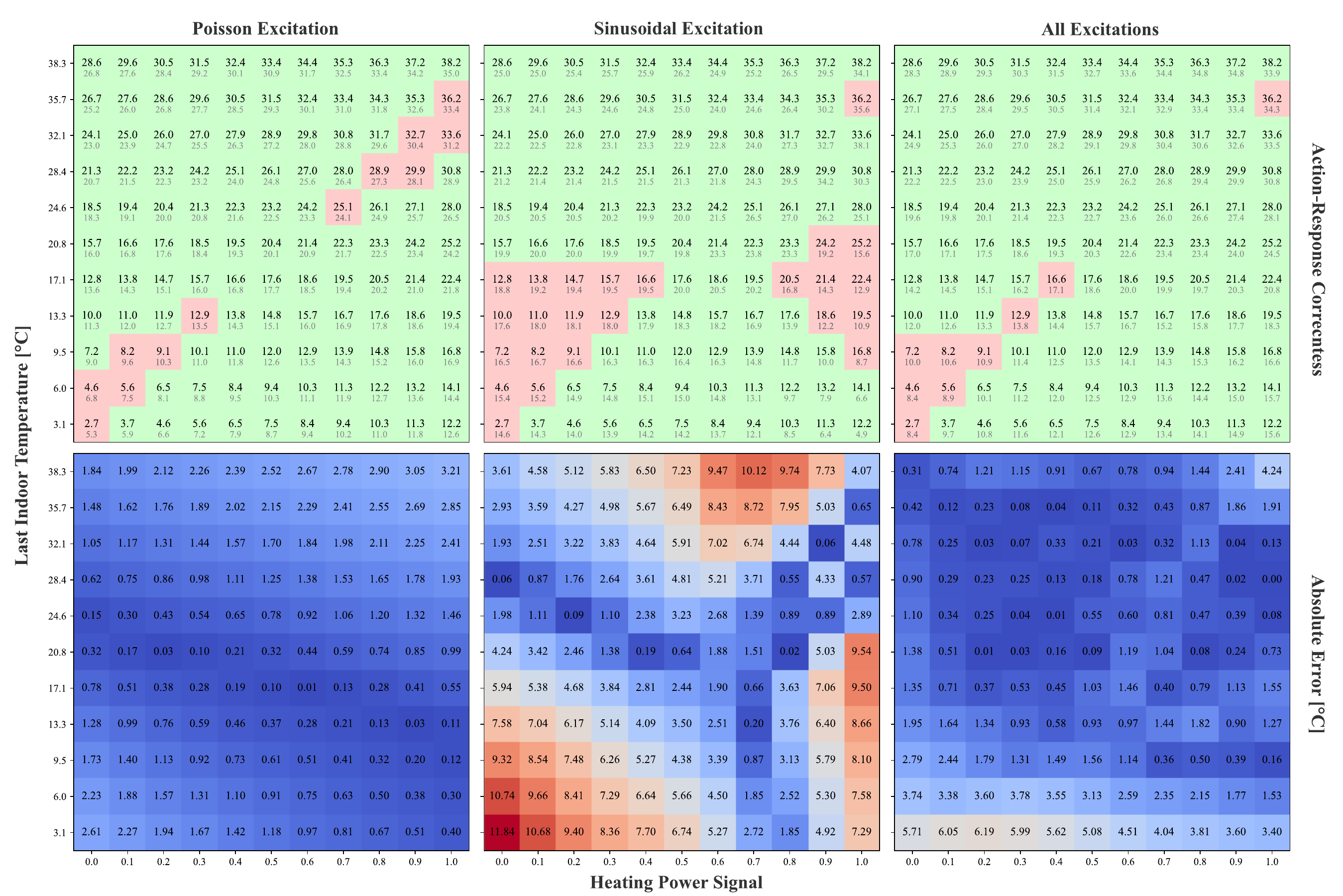}
  \caption{Complementary Absolute Error and Action-Response Correctness Heatmaps for other excitation techniques. The "All Excitations" strategy includes all three excitation techniques during training data generation.}
  \label{fig:combined_eval_heatmap_app}
\end{figure*}

% Ansonsten kann es sein, dass Bilder nicht richtig geladen werden. Soll dann für die Abgabe entfernt werden. 
%\clearpage

\end{document}